\documentclass[aps,twocolumn,showpacs,amsmath,final,amssymb]{revtex4-1}
\usepackage{float}
\usepackage{amsmath}
\usepackage[dvips,final]{graphicx}
\usepackage{epsfig}
\usepackage{color}
\begin{document}

\title{Electronic and magnetic properties of bimetallic L1$_0$ cuboctahedral 
clusters by means of a fully relativistic density functional based calculations}

\author{R. Cuadrado}
\affiliation{Department of Physics, University of York, York YO10 5DD,
             United Kingdom}

\author{R. W. Chantrell}
\affiliation{Department of Physics, University of York, York YO10 5DD,
             United Kingdom}

\date{\today}

\begin{abstract}
By means of density functional theory~(DFT) and the generalized
gradient approximation~(GGA) we present a structural, electronic
and magnetic study of FePt, CoPt, FeAu and FePd based L1$_0$
ordered cuboctahedral nanoparticles, with total numbers of atoms,
N$_{tot}$ = 13, 55, 147. After a conjugate gradient relaxation,
the nanoparticles retain their L1$_0$ symmetry, but the small
displacements of the atomic positions tune the electronic and
magnetic properties. The value of the total magnetic moment
stabilizes as the size increases. We also show that the
Magnetic Anisotropy Energy~(MAE) depends on the size as well as the
position of the Fe-atomic planes in the clusters. We address
the influence on the MAE of the surface shape, finding a small
in-plane MAE for (Fe,Co)$_{24}$Pt$_{31}$ nanoparticles.
\end{abstract}


\maketitle

\section{Introduction}\label{introduction-sec}
Current  uses of nanometer designed magnetic devices range from 
biomedical applications~\cite{bio1,bio2,bio3}, catalysis~\cite{zyade}, 
energy harvesting~\cite{harvesting1} to data storage~\cite{gutf,terris,
10Tbit2,xu}. Because of these diverse applications in nanotechnology
\cite{moser,zeng}, the development of nanostructured magnetic materials 
has become a highly active field. Focusing in the field of magnetic 
recording, a huge amount of experimental and theoretical work has been 
carried out during the last decade to seek novel approaches to construct 
advanced materials for ultrahigh density magnetic storage, with the aim 
of increasing the state-of-the-art beyond 1~Tbit/in$^{2(}$~\cite{1-2Tbit1,
1-2Tbit2,10Tbit1,10Tbit2,10Tbit3}$^)$. Most approaches are focused on 
thin films or multilayers~\cite{thin1,multi1,multi2} and recently on 
slabs~\cite{Chepulskii2012}. However, during the last decade has emerged 
the possibility to use clusters deposited on surfaces~\cite{gambardella,
yan,medina,tamada,zhou,thomson,sebastian2012} to increase the recording 
density. These clusters or nanoparticles~(NPs) have properties different 
from those of bulk alloys due to their reduced surface atomic coordination. 
In particular, binary 3$d$--5$d$ NPs formed by transition metals~(TM) such 
as Fe or Co together with 5$d$ noble metals like Au or Pt allow the 
possibility to tune the magnetic properties based on an in--depth knowledge 
of their geometrical~\cite{rollman,gruner1,singh} and magnetic behavior
\cite{pastor,seivane,sahoo}.

One vital physical quantity in magnetic recording is the magnetic
anisotropy energy~(MAE) of the storage medium. The MAE determines the 
tendency of the magnetization to align along some specific axis in solids 
and clusters. As we have pointed out, binary magnetic NPs based on (Fe,Co)Pt 
are good candidates for novel magnetic recording media, especially those 
phases chemically L1$_0$ ordered, where the value of the MAE is of order 
of 7$\times$10$^7$erg/cm$^{3(}$~\cite{ivanov}$^)$. The trend to higher 
recording densities requires continuous reduction in the grain size while 
retaining large values of the factor $KV/kT$ to avoid loss of recorded 
information due to the onset of superparamagnetic behavior~\cite{super1,
super2}. Following the Neel relaxation law~\cite{neel}, the only way to 
reduce the size of the NPs avoiding this trouble is through a higher values 
of the MAE. To control this magnetic energy, mainly determined by the 
spin--orbit coupling~(SOC)~\cite{stohr}, it is necessary to investigate 
the structure, the electronic and the magnetic behavior of these systems. 
For this purpose, Gambardella~{\it et al} showed experimental and 
theoretically that when Co adatoms were deposited onto a Pt(111) surface, 
they had a MAE of 9~meV/at arising from the strong SOC induced by the Pt 
substrate and for a unquenched orbital moments~\cite{gambardella}. In 
addition, they increased the number of Co atoms on the metal surface 
forming NPs that ranged from 3 up to 40 atoms. The results showed that 
smaller NPs exhibited a higher MAE. These results opened a route to
understand and fabricate high density magnetic recording materials
using deposited NPs on surfaces. There are several experimental
\cite{wang,kang} and theoretical
\cite{seivane,pastor,singh,hong,zhou2,gruner1,gruner2,gruner3,
gruner4,sahoo,antoniak} studies regarding isolated NPs aiming to
obtain the best morphologies and magnetic behavior
covering monometallic NPs~\cite{seivane,pastor,singh},
binary alloys~\cite{gruner1,gruner2,gruner3,gruner4} and even
capped NPs~\cite{sahoo,antoniak}. Gruner~{\it et al} have carried out
a total energy study of a wide range of structures of various
shapes and sizes for (Fe,Co)Pt NPs~\cite{gruner1} as well as for
Fe(Pd,Ni)~\cite{gruner4}.It was found that the most energetically
favored structures obtained were those of ordered multiply twinned
icosahedra and decahedra shapes. Gruner {\it et al} have also obtained
locally the magnetic moment~(MM) for Fe$_{256}$B$_{296}$, with B =
Ni, Pd, Pt, Ir, Au, and as we will see in the present work, the
tendency to augment the Fe MM in the vicinity of the cluster
surface obtained by Gruner is in good agreement with our results.

To obtain the MAE using the framework of DFT implies a huge
computational resource since a fully relativistic~(FR) and
a full potential~(FP) treatment becomes necessary. A widely used
approximation to overcome the all--electron~(full potential)
problem and to do quick and accurate calculations is substitute
the core electrons by a pseudopotential~(PP)~\cite{kleinman,chiang}.
Most of the codes that use the PP approximation use the scalar-
relativistic~(SR) corrections~(Darwin term and mass velocity)
but they are not sufficient to address the behavior of
magnetic systems because  the MAE is mostly controlled by the SOC.
Also, the magnetization density vector can vary from point to
point in space presenting a spin non collinearity. To overcome this
barrier we have used a fully relativistic pseudopotential~(FR-PP)
approach implemented recently in the SIESTA code~\cite{siesta,LS-paper}.

The (Fe,Co)Pt L1$_0$ based alloys have large uniaxial anisotropy
because of the layered structure~(see Fig.~\ref{fig-NPs}).
The purpose of this paper is to investigate the structural,
electronic and magnetic properties of (Fe,Co)Pt and Fe(Au,Pd)
L1$_0$ cuboctahedral nanostructured NPs having the total number of
atoms N$_{tot}$ = 13, 55 and 147, and to calculate the MAE using the 
above mentioned FR-PP scheme. It is shown that the energy surface can 
become complex, showing features beyond the simple uniaxial anisotropy. 
This demonstrates the importance of investigating the dependence of the 
total energy on the orientation of the magnetization axis  as we will 
see in Sec.~\ref{MAE-subsec}.

The paper is structured as follows. In section~\ref{tools-sec}
we describe briefly the theoretical tools to perform all the
calculations as well as the kind of NPs studied in the present
work. The importance of the structural relaxations will be
explained in~\ref{relax-subsec}. The local magnetic moments and the 
density of states are described in subsections~\ref{mm-subsec}
and~\ref{dos-subsec}, respectively. The MAE and its separate
contributions is discussed in Sec.~\ref{MAE-subsec}. Finally,
Sec.~\ref{conclusions-sec} summarizes the main results.

\section{Computational details}\label{tools-sec}
We have undertaken calculations of electronic structure and
magnetic anisotropy energies~(MAE) by means of DFT using a recent
fully relativistic~(FR) implementation~\cite{LS-paper} in the
GREEN~\cite{green,greenp} code employing the SIESTA~\cite{siesta}
framework. We use fully separable Kleinmann-Bylander~\cite{kb}
and norm-conserving pseudopotentials~(PP) of the
Troulliers-Martins~\cite{tm} type to describe the core electrons.
Our DFT based calculations have been performed within the generalized
gradient approximation~(GGA) for the exchange correlation~(XC)
potential following the Perdew, Burke, and Ernzerhof~(PBE)
version~\cite{pbe}. To address the description of magnetic systems, 
pseudocore (pc) corrections were used to include in the XC terms not only 
the valence charge density but also the core charge as Louie~{\it et al}
\cite{cc} pointed out. In order to ease the convergence of three center 
integrals with the size of the real space grid, $\rho^c(r)$ is replaced 
by a pseudo-core charge density, $\rho^{pc}(r)$, which equals the real 
core charge density beyond a given radius, $r_{pc}$, while close to the 
nuclei it becomes a smooth function. The radius $r_{pc}$ should be chosen 
small enough to ensure that the overlap region between the valence and
the core charges is fully taken into account. Based on previous studies
of the binary alloys~\cite{LS-paper}, we have choosen for the radius
that equals the core and valence charge the values of r$_{pc}$(Fe,Co)
= 0.6~Bohrs and r$_{pc}$(Pt,Au,Pd) = 1.0~Bohrs, ensuring that the
overlap region between the valence and the core charge is fully take
into account. As basis set, we have employed double-zeta polarized
(DZP) strictly localized numerical atomic orbitals~(AO). The
confinement energy, $E_c$, defined as the energy cost to confine
the wave function within a given radius was set to 100~meV. The
Fermi-Dirac distribution was used to obtain the ocupation numbers
and the electronic temperature was set to 50~meV.

\begin{figure}[tb]
 \includegraphics[scale=0.32]{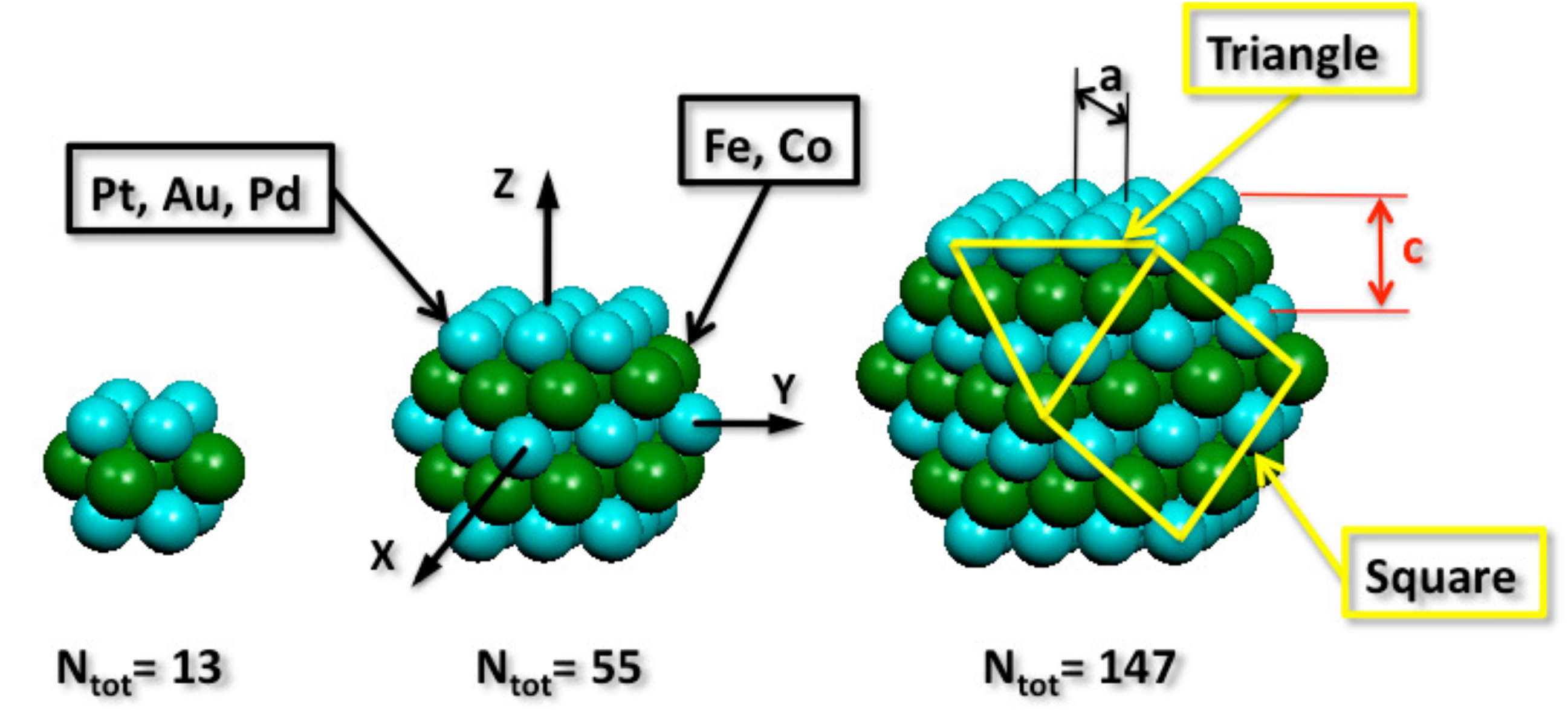}
 \caption{(Color online) Initial L1$_0$ cuboctahedral structures employed 
          in the simulations. Green spheres depict the magnetic atoms
          while turquoise ones are the non magnetic species. On the
          NP at the extreme right, the yellow lines show two kinds of
          surfaces~(square and triangle). The three axis in the
          second NP~(N$_{tot}$=55) represent the usual cartesian frame,
          X, Y, Z being the angles $\theta$ and $\phi$ for each one of 
          them (0$^\circ$,0$^\circ$), (90$^\circ$,0$^\circ$) and
          (90$^\circ$,90$^\circ$), respectively.
          The distance between planes are marked in red by $c$
          parameter and the lattice constant by $a$.}
 \label{fig-NPs}
\end{figure}

In the SR pseudopotential approximation, the Kohn-Sham Hamiltonian
\begin{equation}\label{hamil_1}
\hat H^{KS} = \hat T + \hat V^{local}+ \hat V^{KB} + \hat V_H + \hat V_{XC}
\end{equation}
is diagonal in spin space and collinear spin is assumed~\cite{hedin}.
In equation~(\ref{hamil_1}) $\hat T$ is the kinetic energy, $\hat V^{local}$
is the fully local long-ranged potential commonly set to the $l=0$
radial component of the PP, $\hat V^{KB}$ is the Kleinmann-Bylander~(KB)
operator~\cite{kb}, $\hat V_H$ the Hartree term and $\hat V_{XC}$ is
the exchange-correlation operator. Just two of those terms depend on
the spin projections~(say along the $z$ axis),
$\sigma$~(=$\uparrow,\downarrow$):
the KB term and the final (exchange correlation) term. In the collinear 
case, there is a common quantization axis for the whole system, and the 
charge density has two independent projections,
$\rho^\uparrow({\bf r})$ and $\rho^\downarrow({\bf r})$, parallel and
antiparallel, respectively. However, in the FR-PP aproximation, off-diagonal
spin terms appear in the Hamiltonian causing a mixture of spin components
because the spin quantization axis varies from point to point in space --i.e.
non collinear case. Consequently it was necessary to use the scheme developed 
by K\"ubler~{\it et al}~\cite{kubler} that will give the mixed components for
XC potential~(see Ref.~\cite{LS-paper} for details).

From an {\it ab initio} point of view the MAE is defined as the difference
in the total energy between easy and hard magnetization axis. It is common
to fix the spin quantization axis as the $z$ direction. However, when the
FR-PP approximation is used and we need the total energy in several directions
it necessary to proceed in a different way, specifically to generalize the
magnetization direction to an arbitrary axis, $S_{\bf u}$, characterized by 
polar angles $\theta$ and $\phi$. The procedure will give us a new set of
matrix elements as a function of $\theta$ and $\phi$ angles for the KB
term, $\hat V^{KB}_{\theta,\phi}$. For the total energy calculations
required to determine the MAE we obtain self-consistency by means of the
Hamiltonian instead of using the density matrix. To this end, in each 
iteration the Hamiltonian is obtained after a Pulay mixing~\cite{pulay} 
of the input and output Hamiltonian, $H^{in}$, $H^{out}$ respectively.  
The criterion for a self-consistent solution  is the requirement that 
input and output values differ by less than 1~meV. For each different 
set of angles, ($\theta'$,$\phi'$), we restart the self-consistent scheme 
using as input Hamiltonian the one output for the previous angles 
($\theta$,$\phi$):
\begin{equation}
H'^{,in} = H^{out} - V^{KB}_{\theta,\phi} + V^{KB}_{\theta',\phi'}.
\end{equation}
where the primes denote the matrices calculated for angles
($\theta'$,$\phi'$).

The unit cell for L1$_0$ metallic based alloys consists of two
$fcc$ cells displaced along the diagonal of the cube. The presence
of two different kinds of atoms generates a vertical distortion so
that its structure is defined by two quantities, the in-plane lattice
parameter, $a$, and the out-of-plane constant, $c$.
Prior to relaxation the NPs were constructed from their bulk {\it fct}
phase forming a perfect L1$_0$ ordered cuboctahedron~(see Fig.~\ref{fig-NPs}).
We have restricted
our study to the so--called magic cluster sizes N$_{tot}$ =
(10$n^3$+15$n^2$+11$n$+3)/3, where $n$ is the number of geometrical
closed shells, being the total number of atoms for each species
N$_{M}$ =(5$n^3$+6$n^2$+4$n$)/3 for magnetic~(M) species -Fe and Co-,
and N$_{NM}$ =(5$n^3$+9$n^2$+7$n$+3)/3 for non-magnetic~(NM) species
-Pt, Au and Pd. The initial lattice parameters, $a$, as well as the
$c/a$ ratios were chosen as their bulk experimental values
\cite{galanakis}: a$_{FePt}$ = 3.86 \AA\ and (c/a)$_{FePt}$ = 0.98;
a$_{FePd}$ = 3.89 \AA\ and (c/a)$_{FePd}$ = 0.938; a$_{FeAu}$ =
4.08 \AA\ and (c/a)$_{FeAu}$ = 0.939; a$_{CoPt}$ = 3.81 \AA\ and
(c/a)$_{CoPt}$ = 0.968.

\section{Results}\label{results-sec}
We have carried out a systematic study of bimetallic nanoclusters,
concentrating on the magnetic properties~(spin and MAE). The calculations
were made on fully relaxed structures produced using a conjugate gradient
method. In the following sections we present the results of the magnetic
property calculations, but first we consider the structures themselves
as a basis for the interpretation of the magnetic properties.
\subsection{Conjugate gradient relaxations}\label{relax-subsec}
To carry out relaxation of the NP structures, we have employed
the conjugate gradient~(CG) method, minimizing the
forces between atoms until they were less than 0.03~eV/\AA.
The optimizations have been done at a spin polarized SR level,
and just to address the calculations of MAE, spin moments and density
of states~(DOS) a FR-PP scheme was included.
\begin{figure}[ht]
 \includegraphics[scale=0.45]{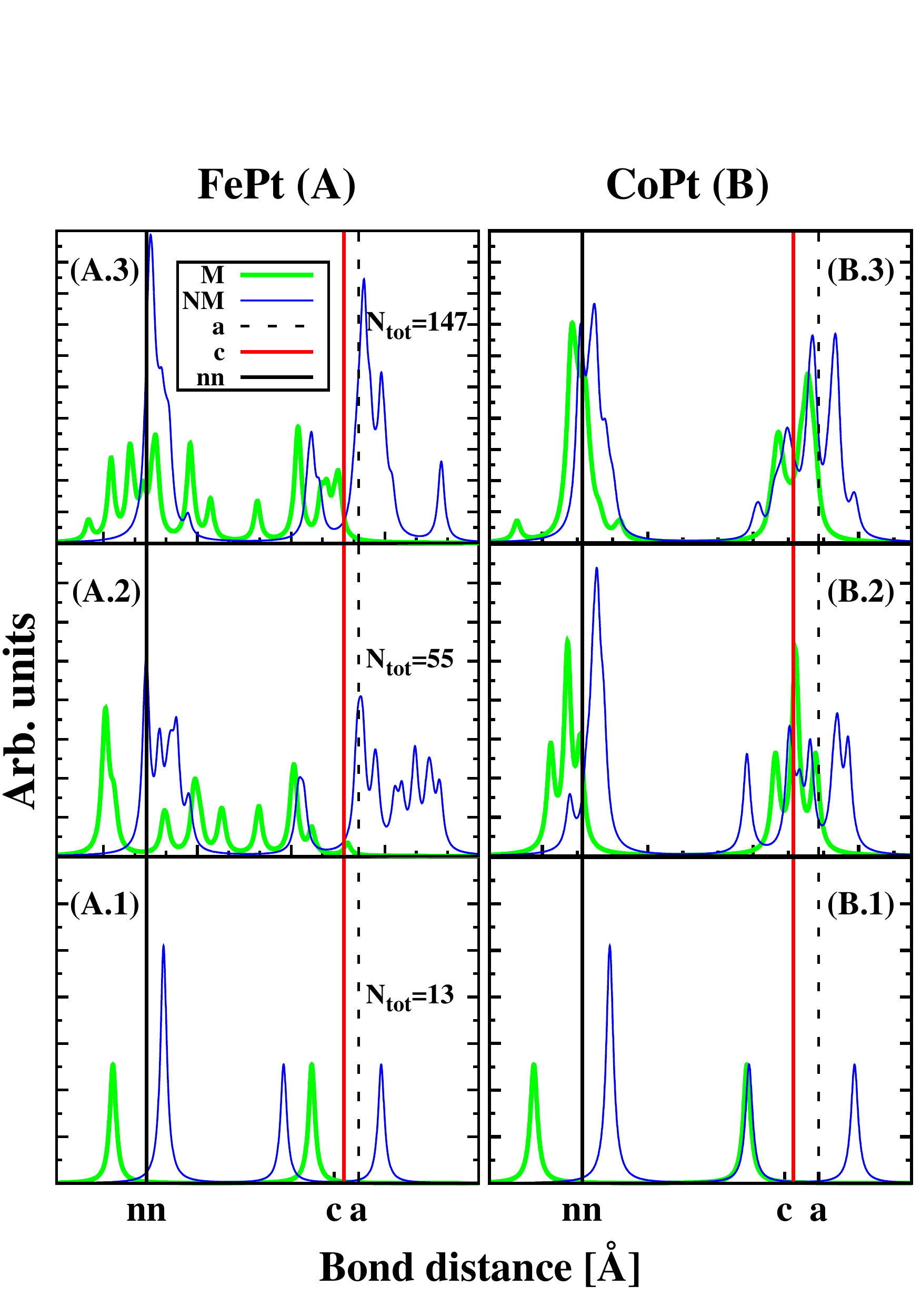}
 \caption{(Color online) Lorentzian broadening of the bond distances 
          between magnetic~(d$_{M-M}$) and non--magnetic~(d$_{NM-NM}$)
          atoms, thick green and thin blue lines, respectively. Each
          one of the two columns~(A,B), from bottom to top,
          depict the distances when the number of total atoms,
          N$_{tot}$, increase from 13 to 147, (A.1) to (A.3) for
          FePt and (B.1) to (B.3) for CoPt. The three vertical
          lines show $a$~(black dashed), $c$~(red solid) and the first 
          nearest neighbours~$nn$~(black solid) experimental lattice 
          values in their bulk phases. The values are provided in the 
          text.}
 \label{dis-fig}
\end{figure}

In Fig.~\ref{dis-fig} we show the evolution of the values of the
lattice parameters $a$, $c$ and the first nearest neighbor
distances, $nn$, after a CG relaxation of the (Fe,Co)Pt NPs.
Although the NPs experience only a small reconstruction,
this is enough to change the magnetic properties substantially
as we will show in section~\ref{MAE-subsec}. The dispersion in
the $nn$ values for Fe species~(green thick lines in the Fig.~\ref{dis-fig}
(A.1--A.4)) around the bulk value is $\pm$0.4 \AA. This means
that the Fe atoms have experienced a regular distribution
around $nn$. Regarding Pt atoms~(blue thin lines), this range
is 0.2 \AA\ less than those of Fe atoms and also the average
displacements are smaller. As a result, the Pt atoms are concentrated
closer to their bulk positions than Fe. With respect to the
mean distance between planes for each species, it is interesting
to note that Fe planes are closer after a reduction of the
bulk value by -0.6 \AA. On the contrary the distances between Pt
planes are larger, increasing by a value of 0.6 \AA. In general,
we can say that the magnetic species has a higher dispersion around
its bulk lattice parameters than the non magnetic one, except for
the case of N$_{tot}$ = 55, where Pt atoms are also significantly
distributed around $c$ and $a$. It can be seen that for CoPt
NPs~(Fig.~\ref{dis-fig} (B.1--B4)) the dispersion around $a$, $c$
and $nn$ is less than for FePt NPs. In this case, both Co and Pt
atom positions deviate by $\pm$0.15 \AA\ from their bulk $nn$
structure values. As in the FePt case the Co atoms have reduced
their mean separation values whilst those of the Pt atoms have 
increased. The distance between planes differ by smaller amounts than 
for FePt, the ranges being between -0.2 \AA\ and +0.3 \AA\ for M and 
NM atoms. This implies that for CoPt NPs there is less distortion of 
the bulk structure. The bond distances for Fe(Au,Pd) relaxed
NPs~(not shown here) have a similar behavior for each atomic
species~(magnetic and non magnetic). Specifically, the Au
atoms experience an increase in their $nn$ distances of 0.15 \AA\
while the separation of Fe atoms decreases by 0.3 \AA. The out of
plane variations are between $+$0.4 \AA\ for Au atoms and $-$0.3 \AA\
for Fe atoms. It is interesting to point out that in general the
distance of the surface atoms from the center of the NPs tends to
be reduced in comparison with the initial bulk structures. Studying 
this distances for the atoms located at different type of surfaces
(squares on the top and the bottom and triangles or squares in the 
side of the molecules)~(See Fig.~\ref{fig-NPs}), we can say that 
there is not a general trend either for magnetic nor non-magnetic 
species.

\subsection{Magnetic moments}\label{mm-subsec}
\begin{figure}[bt]
  \begin{center}
  \includegraphics[scale=0.36,angle=-90]{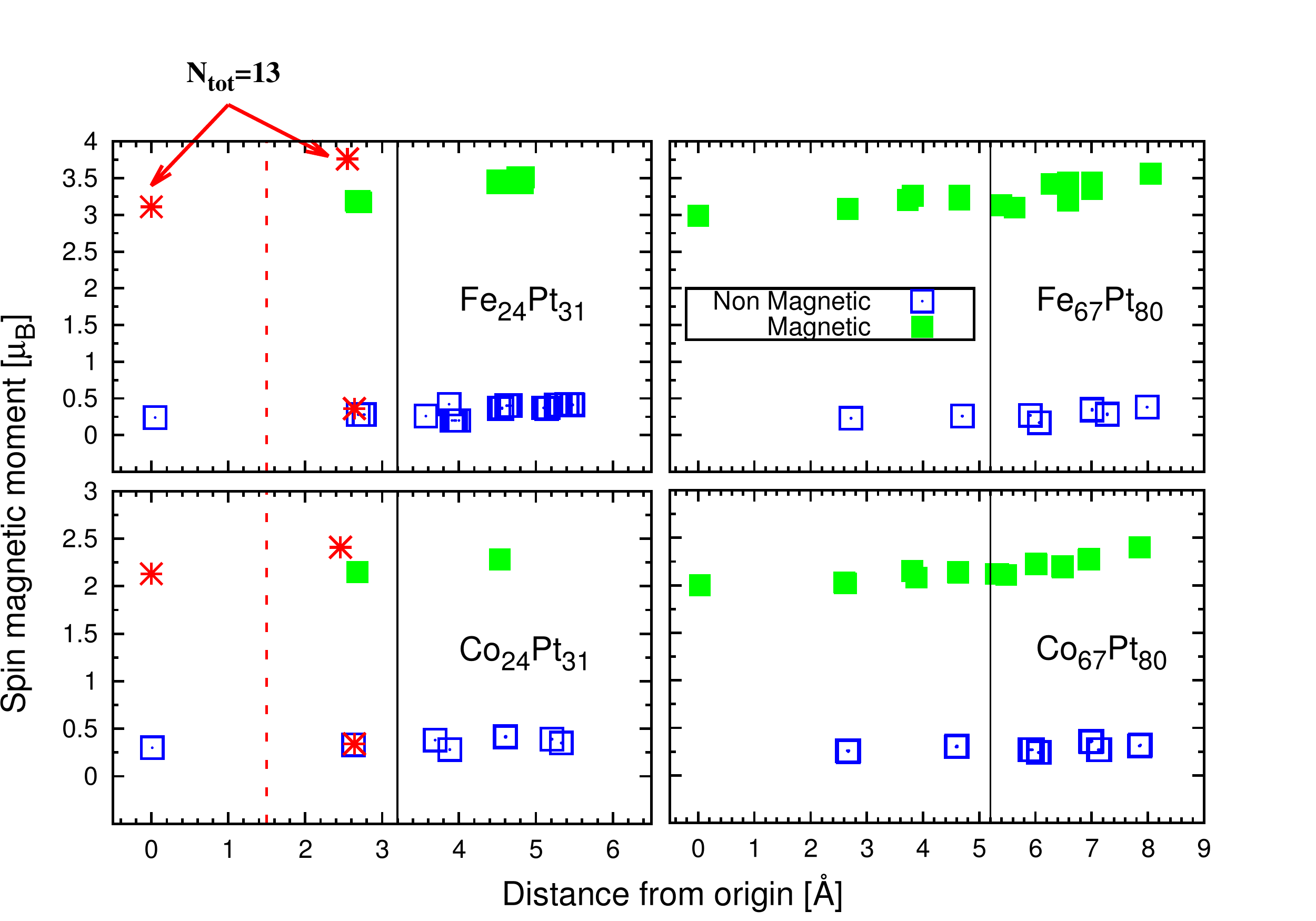}
  \end{center}
  \caption{(Color online) Spin magnetic moments for each kind of
           atom~(magnetic and non-magnetic) as a function of their
           distance from the center of the NP. Each row show,
           from left to right, how the MM values evolve when the
           total number of atoms increase for a specific alloy~(N$_{tot}$
           = 13, 55, 147). Fe and Co atoms are depicted by green filled 
           squares while Pt by empty blue ones. The red symbols show the 
           MM values for each one of the atoms when N$_{tot}$ = 13. 
           Every graph has been divided by a solid line showing two 
           zones that point out the core and surface regions. The added 
           red dashed line in the first column also divide the smaller 
           NP in the core and surface regions.}
  \label{MMs-fig}
\end{figure}

As a result of a Mulliken analysis we show in Fig.~\ref{MMs-fig}
the variation of the spin magnetic moment~(MM) values for every atom
belonging to (Fe,Co)Pt NPs in function of its distance from the center.
The MM values have been derived as the difference between the majority
and minority spin charges and, although this is a qualitative study, it
allows us to determine whether the NPs have more polarization in the
surface or in the core.

The local MM of the Fe and Co atoms~(green filled squares), are remarkably 
large in comparison with those of the Pt atoms~(blue empty squares). This 
is a general trend in all the clusters and the average differences range 
from 1.8 to 3.3~$\mu_B$ for Co$_{67}$Pt$_{80}$ and Fe$_{67}$Pt$_{80}$,
respectively. Taking into account the region where the atoms reside (core 
or surface), the values will be slightly different. So, we can observe 
that for magnetic atoms all the NPs have their inner MM values lower 
than those in the surface by $\thicksim$0.4~$\mu_B$. However, the Pt 
MM values remain around 0.25~$\mu_B$ along their radial positions, the 
difference being only 0.1~$\mu_B$ from inner to surface. This behavior 
prevails even for CoPt NPs.
\begin{table}[b]
\begin{tabular}{cccccccc} \hline \hline
       & N$_{tot}$ & &    M     &  &   NM     & &{\bf Total} \\ \hline \hline
 FePt  &{\bf 13}   & &   4.21   &  & 2.63     & & {\bf 1.62} \\
       &{\bf 55}   & &   3.81   &  & 2.95     & & {\bf 1.66} \\
       &{\bf 147}  & &   3.63   &  & 3.04     & & {\bf 1.65} \\ \hline
 FePd  &{\bf 13}   & &   3.98   &  & 2.45     & & {\bf 1.53} \\
       &{\bf 55}   & &   3.66   &  & 2.84     & & {\bf 1.60} \\
       &{\bf 147}  & &   3.59   &  & 3.00     & & {\bf 1.45} \\ \hline
 FeAu  &{\bf 13}   & &   3.60   &  & 2.24     & & {\bf 1.38} \\
       &{\bf 55}   & &   3.21   &  & 2.48     & & {\bf 1.40} \\
       &{\bf 147}  & &   3.17   &  & 2.66     & & {\bf 1.45} \\ \hline
 CoPt  &{\bf 13}   & &   2.90   &  & 1.81     & & {\bf 1.11} \\
       &{\bf 55}   & &   2.71   &  & 2.10     & & {\bf 1.18} \\
       &{\bf 147}  & &   2.54   &  & 2.13     & & {\bf 1.16} \\ \hline \hline
\end{tabular}
\caption{\label{mm-tab}
          Magnetic moment~(MM) values in $\mu_B$/at of all L1$_0$
          cuboctahedral based NPs. The first column displays the
          four kinds of alloyed NPs and the second shows the number
          of total atoms, N$_{tot}$. The MM values in the third and fourth
          column have been calculated by means the equations
          MM$_{tot}$/N$_{M}$ and MM$_{tot}$/N$_{NM}$, respectively,
          where as in the text, M refers to magnetic atoms and NM
          to the non-magnetic ones. The last column show
          MM$_{tot}$/N$_{tot}$.}
\end{table}
It is noticeable that the Co local MM are $\thicksim$1.4~$\mu_B$
smaller than its magnetic counterpart in any other NP,
even for Fe(Au,Pd)~(not shown in Fig.~\ref{MMs-fig}). It is
also interesting to note that the local MM at the surface in the
smaller NPs~(red symbols) have the largest values. The ratio of surface
to volume atoms in these tiny NPs is 12/1 and 92\% of the atoms
are located on the surface. So, the surface effects are more pronounced
at these sizes as we see in the increasing values of the MM.

In table~\ref{mm-tab} we summarize the total MM of all the NPs studied
in this work. One of the main results is that for all the sizes, the
FePt NPs have higher MM$_{tot}$/N$_{tot}$ when compared with any
other kind of NPs. The FePt values range from 1.62~$\mu_B$/at for
Fe$_5$Pt$_8$ to 1.66~$\mu_B$/at for Fe$_{24}$Pt$_{31}$. Despite this
 small increase for MM$_{tot}$/N$_{tot}$, the different kind of NPs follow
the same trend as the NP size increases. Fe$_{67}$Pd$_{80}$
is an exception having a value of 0.15~$\mu_B$/at less than
Fe$_{24}$Pd$_{31}$. If we inspect the third and the fourth columns,
we note that the above mentioned increase of the MM$_{tot}$/N$_{tot}$
is followed by the non-magnetic atoms but that the converse is true for 
the magnetic atoms. This loss of MM$_{tot}$/N$_M$ for the magnetic atoms 
as the size of the NPs increases could be due to the fact that the 
percentage of surface atoms decreases from 80$\%$ with increasing N$_{tot}$ 
and as we have seen in Fig~\ref{MMs-fig} the contribution of the higher 
spin values of the surface atoms will be diminished. The magnetic atoms 
are not entirely responsibility for the overall magnetic behavior, the 
contribution from the spins of non-magnetic atoms is vital to this 
complicated magnetic process.

\subsection{Density of states~(DOS)}\label{dos-subsec}
\begin{figure}[tb]
 \includegraphics[scale=0.45]{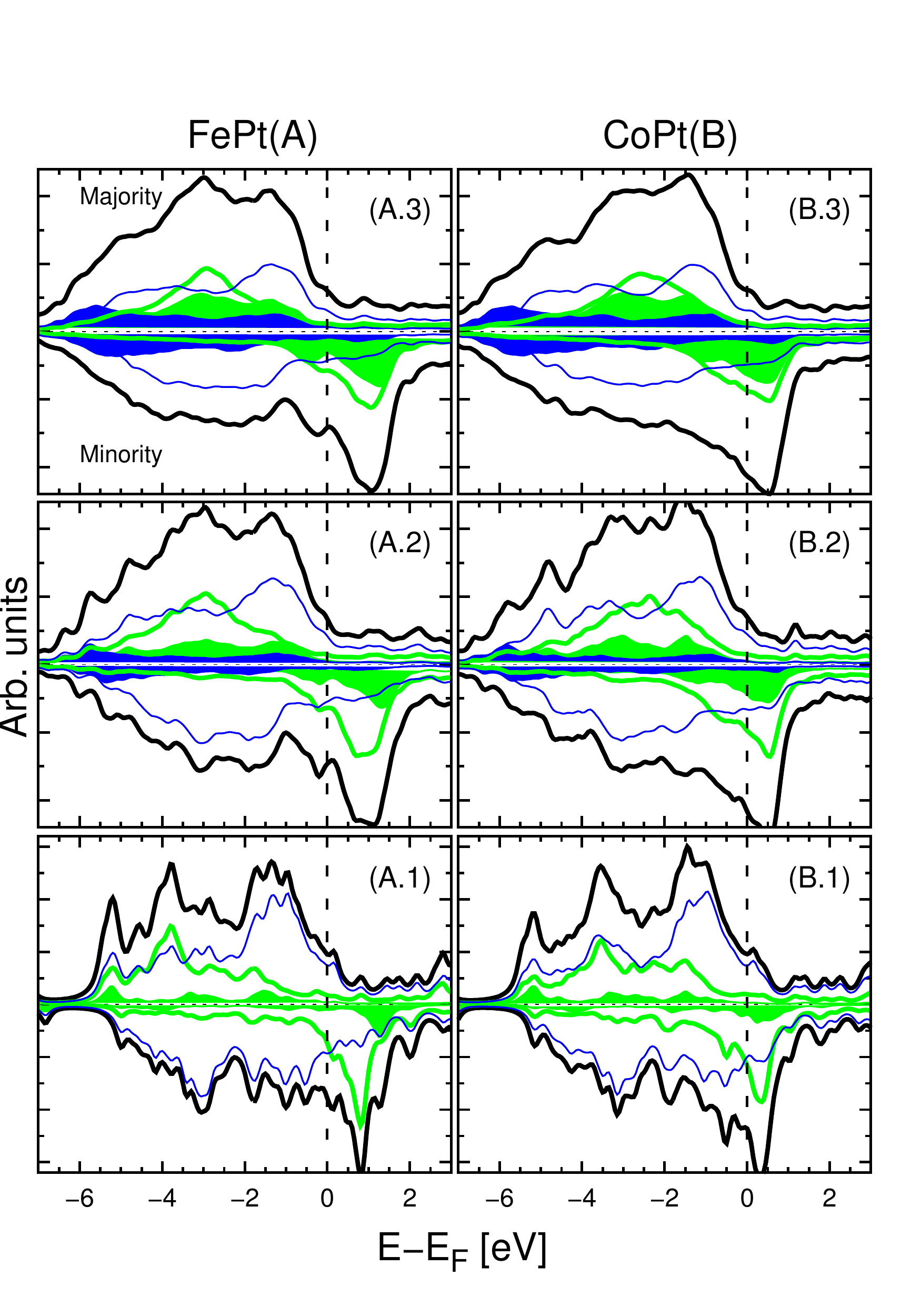}
 \caption{(Color online) Spin resolved density of states~(DOS) of 
          FePt~(A.1-3) and CoPt~(B.1-3) alloyed NPs. Full black 
          lines show the total up and down DOS when the total 
          number of atoms, N$_{tot}$, varies from 13 to 147 from 
          bottom to top. The projected DOS of the surface atoms are 
          represented by thick solid green lines for Fe and Co atoms 
          and thin solid blue lines for the Pt atoms. The filled 
          curves show the projected DOS for the core M and NM atoms, 
          green and blue, respectively.}
 \label{pdos-fig}
\end{figure}

To gain further insight about the electronic behavior of the NPs
we present in Fig.~\ref{pdos-fig} the spin resolved density
of states~(DOS) projected onto Fe, Co and Pt atoms for both FePt
and CoPt NPs, left and right panels, respectively. The atoms have
been divided into two groups as previously in this work:
surface~(thick green lines and thin blue lines) and core~(filled
colored curves). The black lines show the total DOS.

Firstly we note that for both types of NPs, as
N$_{tot}$ increases, the total DOS peaks are smeared and bands are
formed implying that the electrons become delocalized. Further, the
projected DOS on magnetic surface atoms, together with their core
counterpart, show that they provide the largest contribution to the
total MM of the NPs. The majority and minority bands of the smaller
FePt and CoPt type NPs~(A.1,B.1) have peaks around 0.25~eV and as the
size of the NPs increases some of these peaks move below the Fermi
level. For minority states, the FePt peak moves up to -0.3~eV, and
up to the Fermi level for the majority states. CoPt NPs have the same
behavior only for majority states while the minority peak remains
at energies greater than E$_F$ for larger sizes. These displacements 
imply that the $d$ bands are filling and as a result, there is a 
decrease of the total surface MM for the magnetic species of 0.4~$\mu_B$ 
for FePt and 0.15~$\mu_B$ for CoPt~(see first row in the Fig.~\ref{mae-fig})
as N$_{tot}$ increases. Other features of the total majority DOS of 
smaller NPs are the humps at -1~eV, -4~eV and at -5~eV for FePt and at 
-1~eV, -3.5~eV and -5~eV for CoPt. The first two peaks located at higher 
energies persist for larger NPs, however the last disappears when 
N$_{tot}>$13. It is wort noting that the FePt NPs with N$_{tot}>$13 have 
a dip in the minority DOS at the Fermi level, showing that the minority 
channel is dominated by Fe surface atoms. This is in good agreement with 
the work of Gruner~{\it et al.} so that this feature gives a way to
distinguish between different morphologies such as icosahedron or
L1$_0$ cuboctahedron~\cite{rollman}. The DOS of FePd and FeAu
NPs~(not shown here) present a slightly different shape, but nonetheless 
exhibit the main feature the primary responsibility of the M atoms for the
polarization in these NPs.

\subsection{Magnetic anisotropy}\label{MAE-subsec}
\begin{figure}
 \includegraphics[scale=0.45]{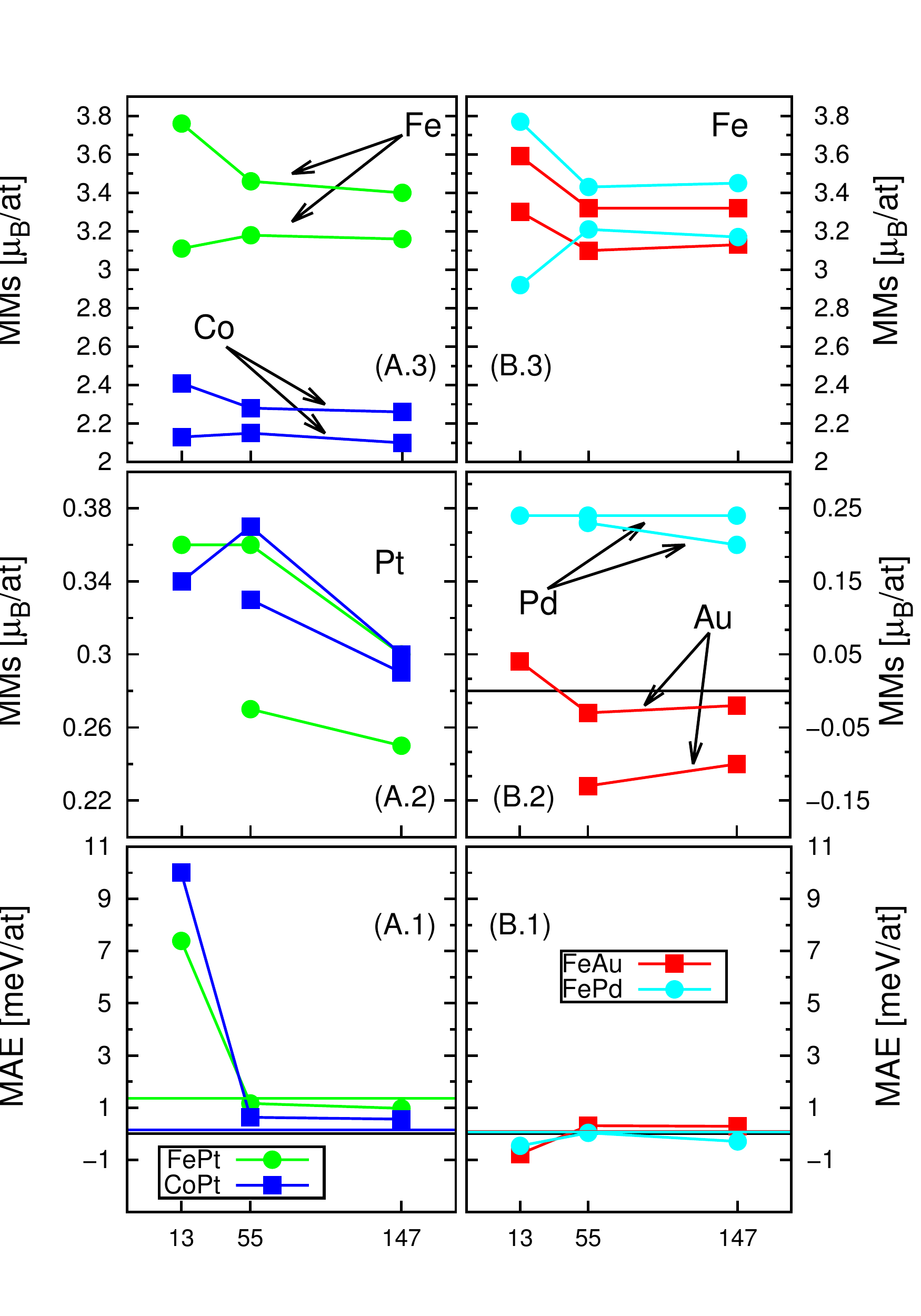}
 \caption{(Color online) Magnetic anisotropy~(MAE) values per atom and 
          mean surface and core magnetic moments~(MM) values per atom 
          for Fe, Co, Au, Pd and Pt of FePt~(green), CoPt~(blue), 
          FeAu~(red) and FePd~(turquoise) NPs as funtion of their total 
          number of atoms.}
 \label{mae-fig}
\end{figure}

We finally present the calculations of the magnetic anisotropy~(MAE).
In order to get a better knowledge of the magnetic behaviour of the NPs,
we also show in Fig.~\ref{mae-fig}, together with the MAE, the MM
for Fe, Co, Pt, Pd and Au atoms in the first and second rows. It is easy
to distinguish between the MM of the surface and core atoms,
whether or not they are magnetic, since as we have seen in sec.~\ref{mm-subsec}
that the local surface MM values are higher than those of the core.

Although there are some common tendencies in the behavior, there is no
overall trend, presumably because of the complexity of the atomic 
rearrangements and charge transfer. Consider first the behavior of the 
MM values of the magnetic atoms in FePt, CoPt, FePd and FeAu. The common 
factor in the behavior of all systems is an increase of the MM in the 
surface over that of the core atoms. In addition, FePt, CoPt and FePd 
exhibit large differences (as large as $\Delta \mu =$ 0.7~$\mu_B$/at for 
FePt), which decreases with increasing N$_{tot}$. The similarity presumably 
reflects the chemical similarity of Pd and Pt. Although its surface atoms 
have a larger MM than the core atoms, FeAu breaks the trend in that 
$\Delta \mu$ remains reasonably constant, presumably reflecting the 
different atomic rearrangements and charge transfer. Turning to the MM 
values of the non magnetic species, the tendency of the MM is to be almost 
constant within both the core and surface regions. Again, we  note that the 
non magnetic atoms of FeAu NPs exhibit  a different trend, and that further 
their MM values for N$_{tot}$ = 55, 147 are negative.

The calculated values of the MAE are shown in Fig.~\ref{mae-fig}(A.1,B.1). 
The alloys from which our NPs have been constructed have in their bulk 
L1$_0$ phases a MAE of order of a few meV~\cite{galanakis} with the easy 
magnetization axis perpendicular to the atomic planes~(see Fig.\ref{fig-NPs}). 
We will see that most of all the studied NPs have the same easy axis 
orientation as their bulk alloys. Also, the values of the total MAE 
obtained in this work are of order of tens of meV following the same 
trend shown by other groups for small NPs~\cite{seivane,sahoo}. In the 
graphs, the MAE is expressed in meV per atom by dividing by the total 
number of atoms (magnetic plus non-magnetic) of each NP and using straight 
coloured lines we show the MAE values for each L1$_0$ alloy.

Consider first the case of FePt and CoPt shown in Fig.~\ref{mae-fig}
(A.1). Although we do not have site-resolved MAE values, we can interpret 
the data in relation to previous calculations of bulk properties of 
FePt~\cite{MAE-T-Staunton,mryasov}. These suggest that the primary 
contribution to the MAE in FePt is a 2-ion anisotropy of the Fe sites 
mediated by the Pt sites. This suggests that the presence of surfaces
and the consequent loss of coordination might be expected to lower the 
overall MAE, which is certainly the case for the two larger NP sizes 
considered here. However, it is interesting to note that the smallest 
NP size exhibits an increased MAE. Although we cannot here obtain 
site-resolved information for the MAE, it seems reasonable to suggest 
that this arises from the modified electronic properties within the 
smallest NPs. This is worth further consideration, with site-resolved 
calculations, since this enhanced MAE may be useful for applications.

\begin{figure}[t]
 \includegraphics[scale=0.37,angle=-90]{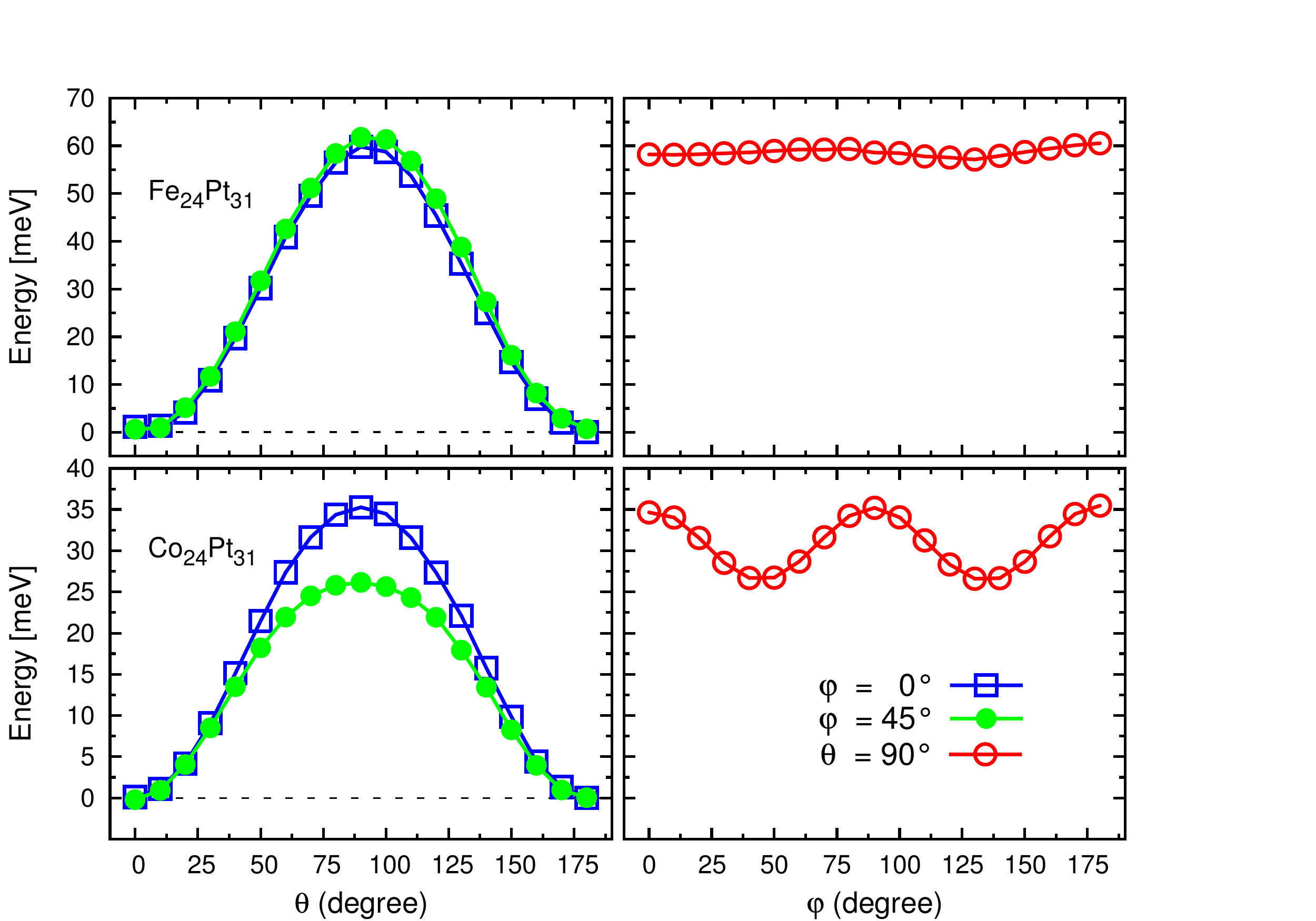}
 \caption{(Color online) Energy variation as a function of the 
          $\theta$(left) and $\phi$(right) angles for Fe$_{24}$Pt$_{31}$ 
          and Co$_{24}$Pt$_{31}$ NPs in the upper and lower rows,
          respectively. The zero of energy is set to the minimum
          value of E and all the points have been joined with lines
          in order to guide the eye.}
 \label{fe+co-mae-fig}
\end{figure}

Regarding the orientation of the easy axis, most of the NPs studied 
present easy axis along the Z--axis. However, we note that for Fe$_5$Pd$_8$, 
Fe$_5$Au$_8$ and Fe$_{67}$Pd$_{80}$ the MAE has a negative value which 
means that the easy axis lies in the XY--plane. Further evidence of 
contributions to the MAE beyond the simple uniaxial case is shown in 
Fig.~\ref{fe+co-mae-fig}. Here we show the variation of the total
energy of the Fe$_{24}$Pt$_{31}$~(upper row) and Co$_{24}$Pt$_{31}$
(lower row) NPs with the $\theta$~(left) and $\phi$~(right) angles.
In both types of NPs the easy magnetization axis lies along the (001)
direction --having the minimum value of the energy~($\theta=0^\circ,
\phi=0^\circ$). Fixing $\phi$ to $0^\circ$~(empty blue squares)
and $45^\circ$~(full green dots) and varying $\theta$ from zero to
180$^\circ$ we obtained different maxima for Co$_{24}$Pt$_{31}$
while the Fe$_{24}$Pt$_{31}$ NPs exhibit purely uniaxial behavior, with 
no dependence of $\phi$. The graphs on the right side sweep the energy 
from $\phi=0^\circ$ to $\phi=180^\circ$ keeping  $\theta$ constant. It 
can be seen that the in-plane magnetization for CoPt has two minima 
exactly at 45$^\circ$ and at 135$^\circ$ degrees~(see Fig.~\ref{fig-NPs}). 
In the case of FePt NPs, no in-plane anisotropy is observed.

\section{Conclusions}\label{conclusions-sec}
In conclusion, we have carried out a fully relativistic calculations, 
within the GGA approximation, of the magnetic moments, density of states 
and MAE of L1$_0$ cuboctahedral FePt, CoPt, FeAu and FePd based NPs. We 
have restricted the total number of atoms to the magic numbers: 13, 55 
and 147, giving diameters of the NPs from 0.6~nm for N$_{tot}$=13 up to 
1.6~nm for N$_{tot}$=147.

Although the original stacking is retained after  CG relaxation, the
atoms exhibit small displacements from their original bulk positions.
The bond distances between magnetic species have higher dispersion 
around the characteristic bulk values than exhibited by non magnetic 
atoms, the non magnetic species being almost at the same position. Although 
this trend is followed by most of the NPs, there is an exception
for CoPt NPs that show much less dispersion for both magnetic and  
non magnetic species.

Regarding the magnetic structure we have shown that the outermost local 
magnetic moments of all the NPs studied in this work are larger than in 
the core, in good agreement with previous investigations. This magnetic 
behaviour is correlated with the PDOS analysis that shows the importance 
of the magnetic ordering of the surface magnetic species polarization. 
Also we showed that the MAE is size and stacking dependent and that the 
value increases for the smallest NPs respect to the bulk values. This 
indicates enhanced thermal stability of the smallest NPs. However, the 
larger FePt and CoPt NPs showed a reduction of the MAE consistent with 
the loss of coordination at the surface and a consequent reduction of 
the (dominant) 2-ion anisotropy. This is an interesting observation which 
shows a dramatic change in the magnetic behavior in the smallest NPs which 
is worth investigating using site-resolved MAE calculations.

The easy magnetization axis generally lies along the (001) direction, 
although in some FeAu and FePd NPs the anisotropy lies in-plane. As an
example of an in-plane magnetic anisotropy we obtained $\theta$ and
$\phi$ energy dependence for (Fe,Co)$_{24}$Pt$_{31}$ NPs in the
Fig.~\ref{fe+co-mae-fig} showing that the surface shape it is important
to study the entire energy surface to investigate the overall form of the 
MAE, which, certainly for the case of Co$_{24}$Pt$_{31}$ has a significant 
contribution from a cubic anisotropy term in addition to the main uniaxial 
term.

\section{Acknowledgments}
The authors would like to acknowledge helpful discussions with Dr. T.J Klemmer. 
The financial support of Seagate Technology and the EU FP7 programme [Grants 
No. NMP3-SL-2008-214469 (UltraMagnetron), No. 214810 (FANTOMAS)] is also 
gratefully acknowledged.

\end{document}